\documentclass[letterpaper]{article}
\usepackage[margin=1in]{geometry}

\title{Controlling Porosity in Supraparticles Composed of Colloidal Rods and Spheres}

\usepackage{authblk}
\author[1, 2]{Kritika Kritika}
\author[3]{Michael P. Howard*}
\author[1, 2]{Arash Nikoubashman*}
\date{*Email: mphoward@auburn.edu, anikouba@ipfdd.de}
\affil[1]{Leibniz-Institut f{\"u}r Polymerforschung Dresden e.V., Hohe Stra{\ss}e 6, 01069 Dresden, Germany}
\affil[2]{Institut f{\"u}r Theoretische Physik, Technische Universit{\"a}t Dresden, 01069 Dresden, Germany}
\affil[3]{Department of Chemical Engineering, Auburn University, Auburn, Alabama 36849, United States}

\usepackage[style=chem-acs]{biblatex}
\DeclareCiteCommand{\citenum}{}{\printfield{labelnumber}}{\multicitedelim}{}
\usepackage{amsmath}

\usepackage{graphicx}
\usepackage{float}

\begin{document}

\maketitle

\begin{abstract}
Supraparticles (SPs) are assemblies of colloidal particles whose properties can be tuned by modifying the chemistry, shape, and size of the colloidal particles as well as their arrangement in the SP. SPs with internal porosity are of particular interest for catalysis, photonics, and adsorption applications because of their high surface area and tunable pore size distribution. SPs are often fabricated by droplet drying, and the nonequilibrium nature of drying processes may provide an additional handle to control particle arrangement within the SP. Here, we use mesoscale particle-based simulations to explore the drying-induced assembly of SPs made from rod-shaped and spherical colloidal particles. We selectively remove one type of particle after drying and characterize the structure of the resulting porous SP. We find that the remaining particles form connected networks for most compositions, with rods percolating at lower volume fractions than spheres. Most of the resulting void volume forms a single contiguous space whose surface area closely follows the total surface area of the remaining component. The pore-size distribution, however, depends strongly on sphere size and on the removed component, reflecting differences in sphere-clustering and rod-bundling before removal. This work provides new insight into how particle size and shape, as well as processing conditions, might be used to manipulate porosity in SPs.
\end{abstract}

\section{Introduction}
Supraparticles (SPs) are mesoscopic structures formed by the organization of colloidal particles into a single larger particle \cite{wendong:2019, wintzheimer:2021, yetkin:2024}. The size, internal morphology, and composition of SPs can be modified through the size, shape, and relative proportion of the constituent colloidal particles \cite{wendong:2019, shim:2021, raju:2021, eren:2022, zhou:2022, liu:jcolloid:2022, yetkin:2024, yetkin:2025, ramachandran:2026}. Many common shapes of colloidal particles cannot fill space completely, so their assembly inevitably produces interstitial voids. This internal porosity is a particularly desirable structural feature for SPs \cite{guo:2022, liu:langmuir:2022} because interconnected voids and high accessible surface area enhance mass transport and interfacial activity for applications such as catalysis, adsorption, sensing, drug delivery, and energy-related technologies \cite{wendong:2019}.

Typically, SPs are formed through droplet-drying processes \cite{wendong:2019, liu:acs:2019, kim:2021, shim:2021, yetkin:2025, yetkin:2024, wang:2024}. As the solvent evaporates, the droplet shrinks and capillary forces bring the colloidal particles into close contact. When evaporation is faster than the characteristic diffusion time of the colloidal particles, their rearrangement into equilibrium mesostructures may be kinetically hindered, potentially giving rise to disordered, more open mesostructures with higher porosity \cite{alshehri:2016, wendong:2019, yetkin:2024, yetkin:2025}. However, precise control over this process remains challenging because the final mesostructure is governed by the complex interplay between drying kinetics, capillary forces, and interparticle interactions. In this context, sacrificial components can be incorporated into the evaporation-induced assembly and then selectively removed after drying \cite{wendong:2019, sultan:2023, madubuko:2024} to provide an additional handle for manipulating the mesostructure of the SP.

\begin{figure}
    \centering
    \includegraphics[width=\linewidth]{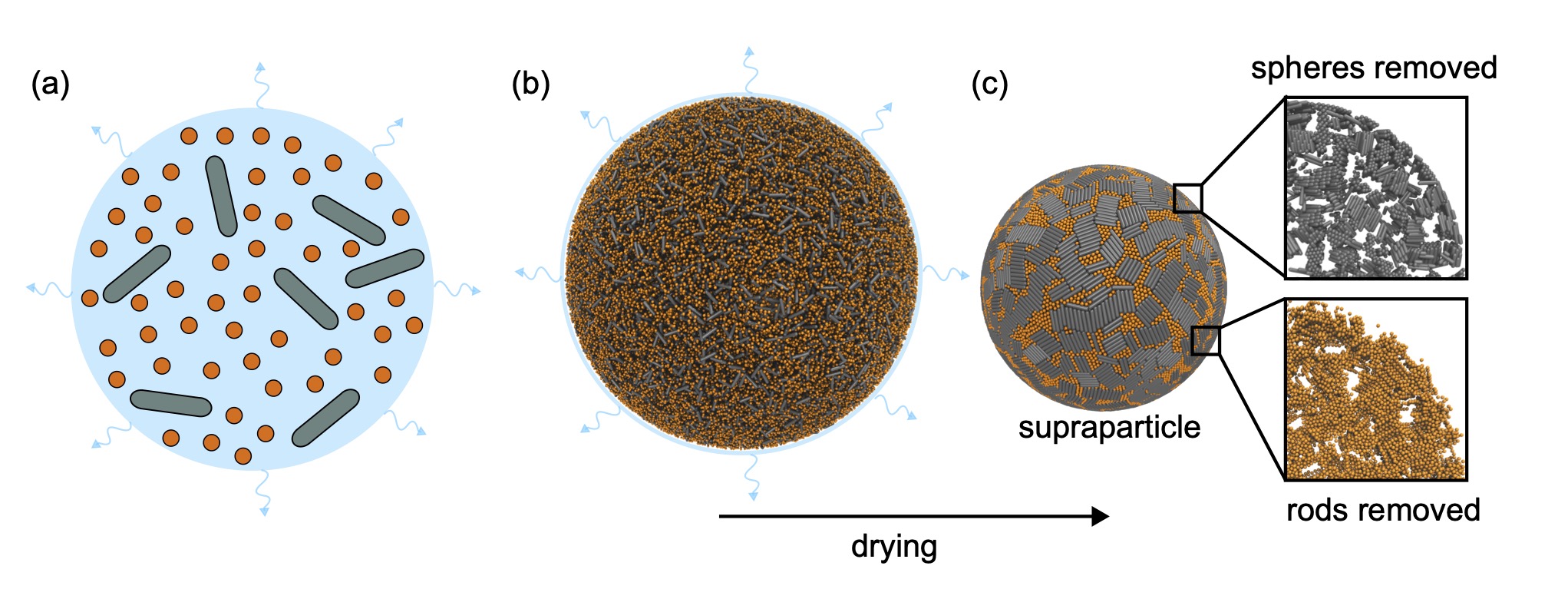}
    \caption{(a) Schematic of a drying droplet containing rod-shaped and spherical colloidal particles. (b) Initial droplet with total volume fraction $\phi_0 = 0.08$. (c) Final SP and the creation of porosity through selective removal of either the spherical or rod-shaped particles, producing interconnected pores. All images of simulations in this article were rendered using Visual Molecular Dynamics (version 1.9.4) \cite{vmd}.}
    \label{fig:schematics}
\end{figure}

Here, we study the fabrication of porous SPs from binary mixtures of rod-shaped and spherical particles with one component selectively removed after the SP is assembled. A schematic of the process is shown in Figure \ref{fig:schematics}, illustrating the initial droplet containing both particle types, the formation of the SP by solvent evaporation, and the subsequent selective removal of either the spherical or the rod-shaped particles to generate a porous SP. We previously showed that SPs containing elongated particles, both on their own and with spherical particles, can exhibit reduced packing efficiency under confinement and during nonequilibrium assembly, leading to enhanced porosity \cite{yetkin:2024}; however, that study was limited to a single size of the spherical particles, a single initial particle composition, and a limited set of drying conditions. It also did not include selective particle removal. Using particle-based simulations, we now systematically examine how the size of the spherical particles, the relative initial volume fractions of rod-shaped and spherical particles, and the drying rate influence the evolving structure of the SP during drying and the pore morphology that results from selectively removing one of the components.

The rest of this article is organized as follows. In Section \ref{section:methods}, we describe the simulation model and methods. In Section \ref{section:results}, we first analyze the structural evolution of the colloidal particles during drying, examining radial volume-fraction profiles to characterize their organization in the SP. We then selectively remove either the spherical or rod-shaped particles, and we characterize the connectivity of the remaining particles and the pores as well as the pore morphology. Finally, in Section \ref{section:conclusions}, we summarize our key findings and discuss their implications for the engineering of porous SPs.

\section{Model and Methods}
\label{section:methods}
\begin{figure}
    \centering
    \includegraphics{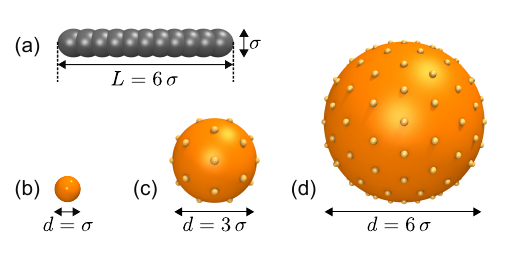}
    \caption{Images of (a) a rod and (b--d) spheres with diameters (b) $1\,\sigma$, (c) $3\,\sigma$, and (d) $6\,\sigma$. The small yellow spheres in (c, d) illustrate the point-like surface sites of the discrete particle models that are coupled to the solvent; they do not exclude any volume.}
    \label{fig.rodsspheres}
\end{figure}

Rod-shaped particles (``rods'') were modeled using overlapping spherical sites (Figure \ref{fig.rodsspheres}a). Each rod consisted of 11 sites of diameter $\sigma$ placed colinearly with a center-to-center separation of $\sigma/2$, resulting in a rod with length $L = 6\,\sigma$ and aspect ratio $\lambda \equiv (L-\sigma)/\sigma = 5$. We considered spherical particles (``spheres'') with three different diameters $d$ relative to the diameter of the rod: $d = 1\,\sigma$, $3\,\sigma$, and $6\,\sigma$ (Figures \ref{fig.rodsspheres}b--\ref{fig.rodsspheres}d). The sphere with $d=1\,\sigma$ was represented using a single central site, while the larger spheres were represented using a discrete particle model consisting of a single central site and several surface sites \cite{poblete:2014, wani:2022}. To obtain the surface sites, we recursively subdivided the faces of a regular icosahedron into equilateral triangles, then projected the resulting vertices onto the surface of a sphere with diameter $d$. A single subdivision step was performed for $d = 3\,\sigma$ to make 42 surface sites, and two subdivision steps were performed for $d = 6\,\sigma$ to make 162 surface sites. This procedure resulted in surface sites that were distributed approximately uniformly on the sphere at a similar surface density ($1.49\,\sigma^{-2}$ and $1.43\,\sigma^{-2}$, respectively). This surface density was found to give good results for several dynamic properties of spheres with diameter $6\,\sigma$ \cite{peng:2024}, and one of us recently showed that these discrete particle models give reasonable results for suspension transport properties when $d=3\,\sigma$ and $6\,\sigma$ \cite{howard:2026}. The sites for a rod and for a sphere with diameter $d=3\,\sigma$ or $6\,\sigma$ were treated as rigid bodies to maintain their geometry \cite{bush:2026}.

Excluded-volume interactions between colloidal particles were modeled using the core-shifted Weeks--Chandler--Andersen (WCA) potential \cite{weeks:1971},
\begin{equation}
U(r) =
\begin{cases}
\displaystyle 4\varepsilon \left[ \left( \frac{\sigma}{r-\Delta} \right)^{12}
- \left( \frac{\sigma}{r-\Delta} \right)^{6} + \frac{1}{4} \right],
& r \le \Delta + 2^{1/6}\sigma \\
0, & {\rm otherwise}
\end{cases},
\label{eq:wca}
\end{equation}
where $r$ is the distance between two interacting sites, $\varepsilon$ sets the strength of the repulsion, and $\Delta$ is the shift parameter that accounts for the particle sizes. The WCA interactions were computed using all sites in the rods and the central sites for the spheres. The shift parameter was $d - \sigma$ for sphere--sphere interactions, $(d - \sigma)/2$ for rod--sphere interactions, and $0$ for rod--rod interactions.

We modeled a freestanding drying droplet as a confining sphere whose radius $R$ decreased over time $t$ as
\begin{equation}
R = \sqrt{R_0^2 - \frac{\alpha}{4\pi} t},
\label{eq:dropletdrying}
\end{equation}
where $R_0$ is the initial droplet radius and $\alpha$ is the rate of change of the droplet's surface area \cite{langmuir:1918}. The interaction between the colloidal particles and the droplet interface was modeled using a purely repulsive harmonic potential that ensured their complete immersion throughout the drying process,
\begin{equation}
U_{\rm d}(r) =
\begin{cases}
\displaystyle \frac{k_{\rm d}}{2} (r - R - \Delta_{\rm d})^2, & r > R + \Delta_{\rm d} \\
0, & {\rm otherwise} \\
\end{cases},
\label{eq:surface:harmonicpotential}
\end{equation}
where $r$ denotes the distance of a site from the center of the droplet, $k_{\rm d}=200\,\varepsilon/\sigma^2$ sets the strength of the repulsion at the interface, and $\Delta_{\rm d}$ is a shift parameter to account for the different particle sizes. This potential was applied to all the sites in the rods and the central sites of the spheres, and the shift parameter was $-\sigma/2$ for the rods and $-d/2$ for the spheres accordingly. Note that modeling the droplet interface using this potential assumes that the droplet retains a spherical shape throughout the drying process and therefore does not allow interfacial deformations such as buckling \cite{zhou:2022, yetkin:2024, yetkin:2025, roemling:2025, mahato:2025}.

Solvent-mediated hydrodynamic interactions can play an important role in nonequilibrium drying processes \cite{kundu:2025, jee:2026}, so the solvent was modeled using multiparticle collision dynamics (MPCD) \cite{malevanets:1999, gompper:2008, howard:2019}. MPCD represents the solvent as point particles of mass $m$ that undergo alternating streaming and collision steps. During the streaming step, solvent particles moved ballistically and freely flowed through the droplet interface, hence filling the entire cubic simulation box at a uniform density. We previously demonstrated that this unbounded-solvent approach produced the same distribution of spheres in a drying droplet as no-penetration boundary conditions at the droplet interface while being considerably simpler to implement \cite{kundu:2025}. During the subsequent collision step, solvent particles were grouped into cubic cells of side length $l=1\,\sigma$ and exchanged momentum with other particles in the same cell using the stochastic rotation dynamics rule without angular momentum conservation \cite{malevanets:1999}. Specifically, the velocities of the particles in a cell relative to the cell's mass-averaged velocity were rotated by a fixed angle around an axis randomly chosen from the unit sphere. A cell-level Maxwell–Boltzmann thermostat \cite{huang:2010, huang:2015} was used to maintain a constant temperature $T=1\,\varepsilon/k_{\rm B}$, where $k_{\rm B}$ is the Boltzmann constant. To preserve Galilean invariance, the collision cells were randomly shifted at each collision step along each Cartesian direction by an amount drawn uniformly from $[-l/2, l/2]$ \cite{ihle:2001, ihle:2003}.

We used a solvent number density of $20\,\sigma^{-3}$, a rotation angle of $130^\circ$, and a collision time of $0.1\,\tau$, where $\tau = \sqrt{m\sigma^2/\varepsilon}$ is the unit of time. The solvent density is larger than typical for MPCD to mitigate solvent compressibility artifacts under fast drying conditions \cite{kundu:2025}. These parameters give an estimated dynamic viscosity of $\eta_0 = 18.21\,\varepsilon\tau/\sigma^3$ for the pure solvent \cite{gompper:2008}. The masses of all sites in the colloidal particles were set to $20\,m$ except for the central site of the spheres with $d=3\,\sigma$ and $6\,\sigma$, which had no mass because it was used only for evaluating eq~\ref{eq:wca}. The colloidal particles were coupled to the solvent by including all sites in the rods, the central site for the sphere with $d=1\,\sigma$, and the surface sites for the spheres with $d=3\,\sigma$ and $6\,\sigma$ in the collision step \cite{malevanets:2000, bush:2026}. Between collisions, the dynamics of the colloidal particles were propagated using velocity Verlet integration with a timestep of $0.005\,\tau$. All simulations were performed using HOOMD-blue \cite{anderson:cms:2020, nguyen:2011, howard:2016, howard:2018, howard:cms:2019, glaser:2020} (version 5.4.0) extended with azplugins \cite{azplugins} (version 1.2.0).

Equilibrated colloidal suspensions were prepared in a droplet with radius $R_0 = 100\,\sigma$ for varying initial volume fractions $\phi_{0, i} = N_i v_i/(4 \pi R_0^3/3)$ of each component, where $N_i$ and $v_i$ denote the number and volume of particle type $i$, respectively. The volume of a sphere was $v_{\rm s} = \pi d^3/6$, while the volume of a rod was estimated as that of a spherocylinder, $v_{\rm r} = (\pi\sigma^3/6)(1+ 3\lambda/2)$. The total initial volume fraction was fixed at $\phi_0 = \phi_{0, {\rm r}} + \phi_{0, {\rm s}} = 0.08$ while varying the volume-fraction ratio of spheres to rods as $\phi_{0,{\rm s}}/\phi_{0,{\rm r}} = 0.25$, 0.5, 1, 2, and 4. The corresponding numbers of spheres and rods, $N_{\rm s}$ and $N_{\rm r}$, are provided in Table S1. In the rest of the paper, we will refer to the initial composition based on the amount of one component $\phi_{0,i}$ relative to the total initial volume fraction $\phi_0$. For each composition, initial configurations were generated using Langevin dynamics with translational friction coefficient $1\,m/\tau$ and isotropic rotational friction coefficient $1\,\tau^{-1}$ for both rods and spheres, starting from random nonoverlapping particle placements within a sphere of radius $R_0 = 232\,\sigma$. The radius of the droplet was then decreased to $R_0 = 100\,\sigma$ at a constant rate over a period of $5 \times 10^4\,\tau$, after which the system was further equilibrated for $5 \times 10^5\,\tau$. Three particle configurations were recorded every $5 \times 10^4\,\tau$ from the end of this equilibration period. This procedure was repeated 5 times for each composition using a different seed for the pseudorandom number generator to create a total of 15 statistically independent particle configurations.

The saved configurations were then used to simulate drying at different rates, which we specified using a P{\'e}clet number defined as
\begin{equation}
{\rm Pe} = \frac{R_0 V_0}{D_{0,{\rm r}}},
\label{eq:pecletnumber}
\end{equation}
where $V_0$ is the initial speed of the receding droplet interface, and $D_{0,{\rm r}}$ is the isotropically averaged translational self-diffusion coefficient of a rod at infinite dilution \cite{bruggen:1997},
\begin{equation}
D_{0, {\rm r}} = \frac{k_{\rm B} T}{3 \pi \eta_0 L} \left[ \ln \left( \lambda+1 \right) + 0.316 + \frac{0.582}{\lambda+1} + \frac{0.050}{\left( \lambda+1 \right)^2} \right],
\label{eq:roddiffusion}
\end{equation}
which is $D_{0,{\rm r}} = 2.14 \times 10^{-3}\,\sigma^2/\tau$ for our rods. Drying simulations were performed at P{\'e}clet numbers ${\rm Pe} = 5$, 10, and 50, and the simulations were stopped when the droplet radius reached $R = 51\,\sigma$, corresponding to a final total volume fraction of approximately $0.60$. This final volume fraction was chosen to compare all systems at the same degree of compaction, which represents a dense, dried SP close to random-packing conditions. Note that we defined ${\rm Pe}$ using the self-diffusion coefficient of a rod, but a P{\'e}clet number can also be defined using the self-diffusion coefficient of a sphere; it is also consistently estimated to be greater than 1 for these drying conditions (Table S2). This procedure was repeated 5 times for each of the three drying rates, with each simulation using a different initial configuration. We report the averages over these independent simulations with uncertainties estimated as one standard error of the mean.

\section{Results and Discussion}
\label{section:results}
\subsection{Particle distribution}
To establish a baseline for assessing the effect of drying on the organization of the colloidal particles in the SP, we first characterized their equilibrium structure in droplets of different radii. Specifically, we performed Langevin dynamics simulations with translational friction coefficient $0.1\,m/\tau$ and isotropic rotational friction coefficient $0.1\,\tau^{-1}$, starting from initial configurations described in Section \ref{section:methods}. The droplet radius was then reduced linearly from $R_0 = 100\,\sigma$ to $R=51\,\sigma$ over a period of $5\times10^4\,\tau$, and configurations at selected radii were recorded. With these friction coefficients and this interface speed ($V_0 \approx 10^{-3}\,\sigma/\tau$), diffusive relaxation was fast relative to compression (${\rm Pe} \ll 1$) and the particles remained close to equilibrium. These intermediate configurations were subsequently equilibrated at fixed radius using Langevin dynamics for an additional $5\times10^3\,\tau$. The average volume fraction of component $i$ at radial distance $r$ from the center of the droplet, $\phi_i(r)$, was calculated using the same approach as in Ref.~\citenum{yetkin:2024}. Results are reported for the final equilibrated configurations at each target radius.

\begin{figure*}
    \centering
    \includegraphics[width=\linewidth]{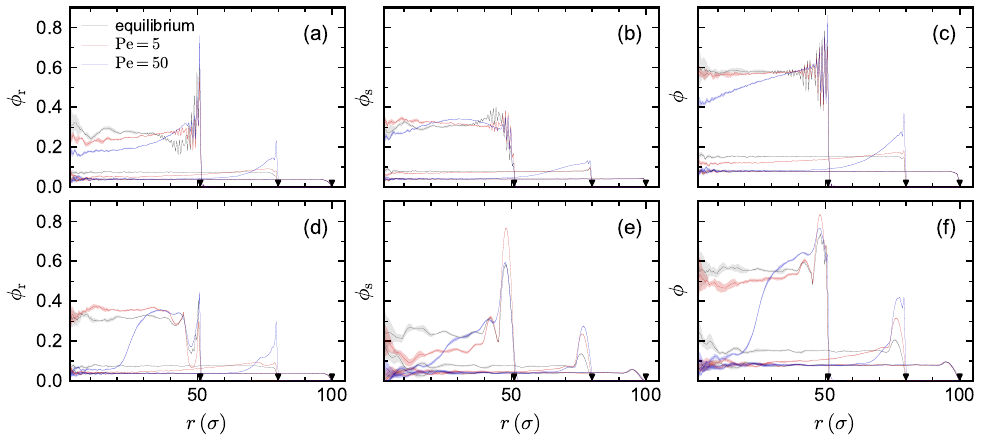}
    \caption{Local volume fraction of (a,d) rods $\phi_{\rm r}$, (b,e) spheres $\phi_{\rm s}$, and (c,f) their sum, $\phi = \phi_{\rm r} + \phi_{\rm s}$ at radial distance $r$ from the center of the droplet for sphere diameters (a--c) $d=1\,\sigma$ and (d--f) $6\,\sigma$ at different stages of drying when $\phi_{0,{\rm s}}/\phi_0=0.5$. Solid lines are profiles in the drying simulations initially ($R = R_0=100\,\sigma$), when the radius was about $R = 79.9\,\sigma$, and at the end ($R=51\,\sigma$) when ${\rm Pe} = 5$ or 50, while dashed lines are the corresponding equilibrium profiles for these droplet radii. The complete set of volume fraction profiles is provided in Figures S1--S3 for all $d$, $\phi_{0,{\rm s}}/\phi_0$, and Pe.}
    \label{fig:volfrac}
\end{figure*}

Initially, the particle distributions were essentially uniform within the droplet interior for all sphere diameters $d$, with only minor differences at the droplet interface. At smaller droplet radii, the equilibrium distributions became increasingly nonuniform. When $d=1\,\sigma$, the rods were weakly enriched at the droplet interface, accompanied by an enrichment of spheres below the surface (Figures \ref{fig:volfrac}a and \ref{fig:volfrac}b); in contrast, when $d=6\,\sigma$, the spheres preferentially organized near the droplet interface (Figures \ref{fig:volfrac}d and \ref{fig:volfrac}e). Spheres with an intermediate diameter of $d=3\,\sigma$ showed similar particle distributions as those with $d=6\,\sigma$ (Figure S2). 

\begin{figure*}
    \centering
    \includegraphics[width=\linewidth]{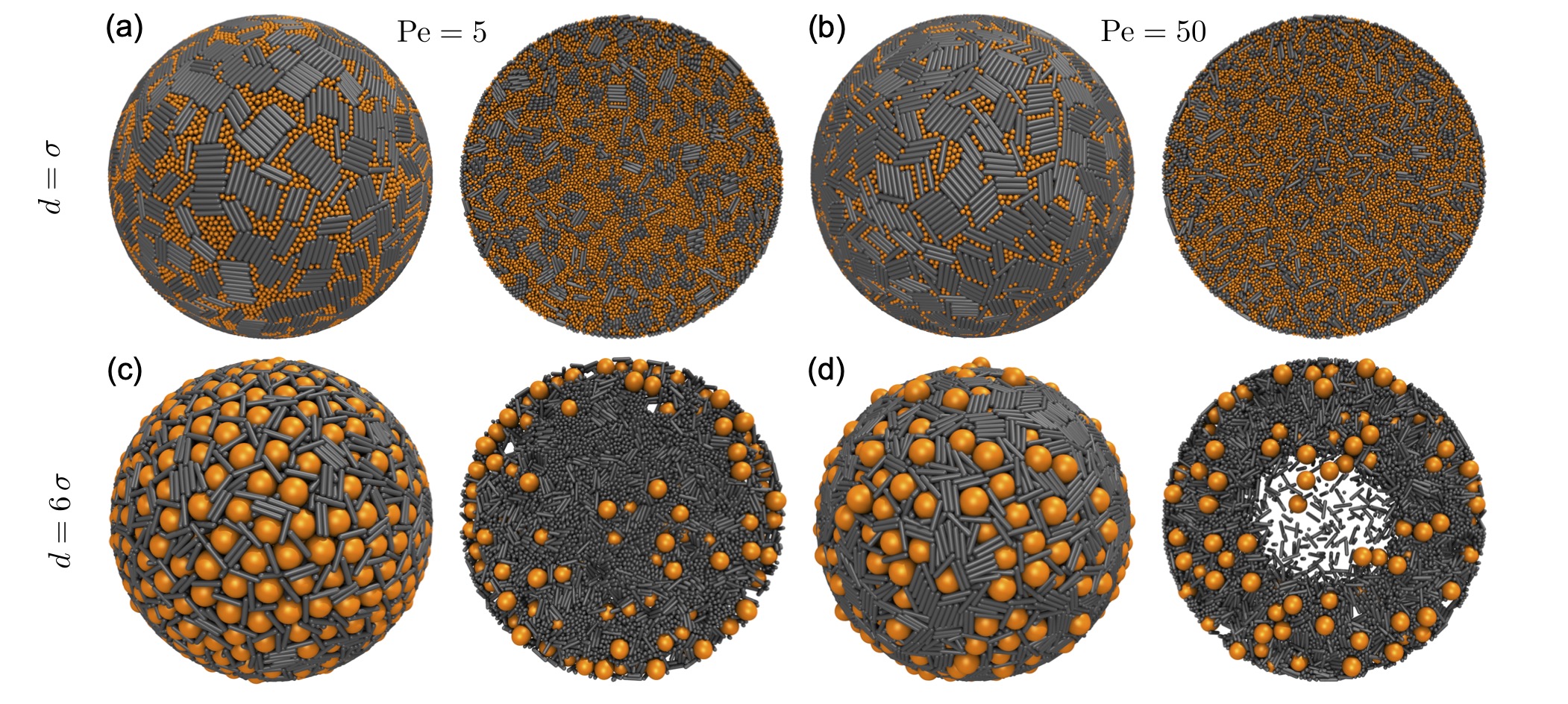}
    \caption{Exterior (left) and cross-sectional (right) views of SPs for $\phi_{0,{\rm s}}/\phi_0 = 0.5$ and (a) $d=1\,\sigma$, ${\rm Pe}=5$; (b) $d=1\,\sigma$, ${\rm Pe}=50$; (c) $d=6\,\sigma$, ${\rm Pe}=5$; and (d) $d=6\,\sigma$, ${\rm Pe}=50$. The complete set of simulation snapshots is provided in Figures S4--S9 for all $d$, $\phi_{0,{\rm s}}/\phi_0$, and Pe.}
    \label{fig:snapshots}
\end{figure*}

We then conducted drying simulations at ${\rm Pe} > 1$ where nonequilibrium structures could form. For spheres with $d=1\,\sigma$, the distribution of particles in the SP closely resembled the equilibrium distribution when dried at the smallest P{\'e}clet number (${\rm Pe} = 5$), with only slightly more rods accumulated at the SP surface. This similarity is reasonable because at this value of ${\rm Pe}$, particle diffusion and advection occur over comparable time scales. Representative snapshots (Figure \ref{fig:snapshots}) further revealed small aligned bundles of rods both near the surface and in the interior of the SP. However, these local bundles did not translate into global nematic order, likely because confinement and kinetic arrest limited long-range alignment. At the largest P{\'e}clet number (${\rm Pe} = 50$), advective transport became more pronounced, leading to stronger accumulation of rods at the SP surface. Despite this pronounced interfacial accumulation of rods, the total volume fraction remained approximately uniform throughout the SP interior (Figure \ref{fig:volfrac}c).

For spheres with $d=6\,\sigma$, a different interfacial organization was observed during drying. Spheres preferentially occupied the SP surface at equilibrium, but with increasing Pe, the rods increasingly accumulated at the SP surface and excluded spheres (Figures \ref{fig:volfrac}d and \ref{fig:volfrac}e). This behavior suggests a form of stratification \cite{howard:2017, fortini:2016}, with rods enriched at the interface and spheres depleted from the interior. The total volume fraction profile further revealed the emergence of a depleted region near the SP center at large Pe (Figure \ref{fig:volfrac}f), consistent with the central void visible in the corresponding snapshots (Figure \ref{fig:snapshots}d). This central void can be attributed to the reduced diffusivity of the spheres with $d=6\,\sigma$ compared to the $d=1\,\sigma$ case, which gives rise to a larger effective Pe for the spheres (Table S2) and thus a stronger influence of advective transport that limits redistribution toward the SP interior. As a result, both rods and spheres accumulated near the boundary, leading to the formation of a low-density region at the center at sufficiently large Pe. Similar behavior has been noted in simulations \cite{kundu:2025, jee:2026} and experiments \cite{fair:2004, bertrand:2005, sloth:2009, lintingre:2016, fu:2017} of drying droplets containing monodisperse and polydisperse particles.

Here, we have shown the particle distributions at equilibrium and during drying for the equal initial composition $\phi_{0,{\rm s}}/\phi_0 = 0.5$ when ${\rm Pe} = 5$ and 50. These distributions were qualitatively similar across all examined initial compositions, and the ${\rm Pe} = 10$ cases showed behavior intermediate between ${\rm Pe} = 5$ and ${\rm Pe}=50$ (Figures S1--S3). These results provide a broader context for our previous simulations \cite{yetkin:2024}, where $d=3\,\sigma$ spheres and $\lambda=6$ rods at equal initial rod and sphere volume fractions produced SPs with a rod-rich shell and a sphere-rich region beneath it. The present results show that this radial segregation is not generic across all sphere sizes and compositions. Rather, changing the sphere diameter relative to the rod thickness can weaken or modify the surface enrichment of rods.

\subsection{Particle connectivity}
After analyzing the distribution of both the spheres and the rods in the dried SP, we next tested the connectivity of the network of particles that remained after selective removal of one component because such connectivity is a key factor in mechanical stability of the SP. To this end, the configuration of particles that remained was discretized on a grid of cubic voxels with edge length $0.1\,\sigma$. Each voxel was assigned a value of 1 (occupied) if its center was inside a remaining particle and 0 (not occupied) otherwise. To reduce artifacts from the grid placement, this procedure was performed on an undisplaced grid and on four additional grids randomly shifted by up to half a voxel in each Cartesian direction, then the five occupancy values were averaged and rounded to either 1 or 0. The connectivity of the discretized remaining particles was characterized using a graph-based approach, where two occupied voxels were considered connected if they were adjacent in the grid along a Cartesian direction. The largest cluster of connected voxels was identified, and we report its size as the fraction $f$ of voxels in the largest cluster relative to the total number of occupied voxels.

\begin{figure*}
    \centering
    \includegraphics[width=\linewidth]{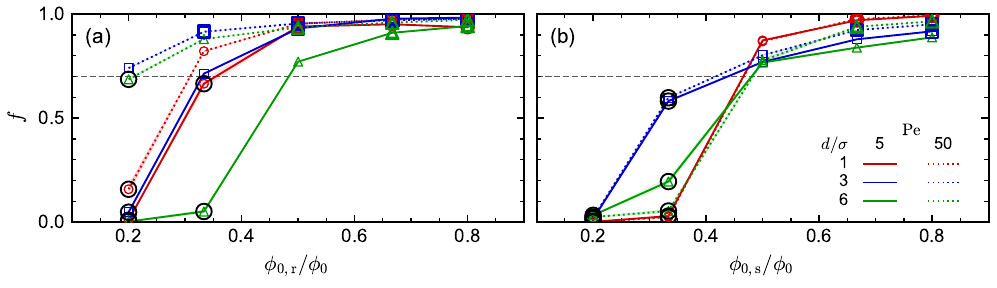}
    \caption{Fraction of occupied voxels $f$ in the largest cluster after removing (a) spheres or (b) rods as a function of initial composition of the particle that remains, (a) rods or (b) spheres, for different sphere diameters $d$ when ${\rm Pe} = 5$ or 50. The dashed horizontal lines show the threshold $0.7$ that we chose to consider the remaining particles percolated. The estimated uncertainty is less than the marker size.}
    \label{fig:particlefraction}
\end{figure*}

We computed $f$ for all cases studied (Figure \ref{fig:particlefraction}), and used $f > 0.7$ as a heuristic to classify the remaining particle network as percolating. Cases with $f \leq 0.7$ are retained in the pore-morphology analysis (Section \ref{section:results:morphology}) but marked in the corresponding figures. When the spheres were removed (Figure \ref{fig:particlefraction}a), the remaining rods formed a percolating network for $\phi_{0,{\rm r}}/\phi_0 > 0.33$ for all cases. At $\phi_{0,{\rm r}}/\phi_0 = 0.33$, the remaining rods formed a system-spanning network for ${\rm Pe}=50$, but not for all lower ${\rm Pe}$ cases. At $\phi_{0,{\rm r}}/\phi_0 = 0.20$, the rods were too sparse to sustain a system-spanning cluster for nearly all sphere diameters and drying rates. On the other hand, when the rods were removed (Figure \ref{fig:particlefraction}b), the remaining spheres formed a percolating network for $\phi_{0,{\rm s}}/\phi_0 \geq 0.5$ across all sphere diameters and drying rates. For $\phi_{0,{\rm s}}/\phi_0<0.5$, the remaining spheres did not form a percolating network for any of the case studied. The fraction of voxels in the largest cluster when ${\rm Pe} = 10$ followed similar trends as when ${\rm Pe}=5$ (Figure S10). Overall, the remaining rods formed connected networks at lower volume fractions than the remaining spheres, making sphere removal a potentially more robust route to fabricating highly porous SPs than rod removal. This behavior is consistent with the lower connectivity threshold of rods, whose elongated shape allows them to bridge larger distances and form percolating networks at lower volume fractions compared to spheres \cite{balberg:1984, schilling:2015}.

\subsection{Pore morphology}
\label{section:results:morphology}
We last investigated the structure of the pores that were formed after selective particle removal. The pore space was identified by inverting the discretized representation of the remaining particles and marking the voxels whose centers lay outside the spherical SP boundary ($R=51\,\sigma$) as not occupied. We then identified the connected voxels and determined the fraction $f_{\rm p}$ of occupied voxels in the largest pore, analogous to $f$. Here, we found $f_{\rm p} \approx 1.0$ for all cases (Figure S11), indicating an almost fully continuous pore network. The total pore volume $V_{\rm p}$ was obtained by summing the volume of all voxels, while the pore surface area $A_{\rm p}$ was calculated from the area of the exposed voxel faces \cite{michielsen:2001, howard:2020}. Because voxelization systematically overestimates surface area, we multiplied $A_{\rm p}$ by the asymptotic correction factor of $2/3$ for discretized smooth surfaces in the limit of vanishing voxel size (Section S5).

\begin{figure*}
    \centering
    \includegraphics[width=\linewidth]{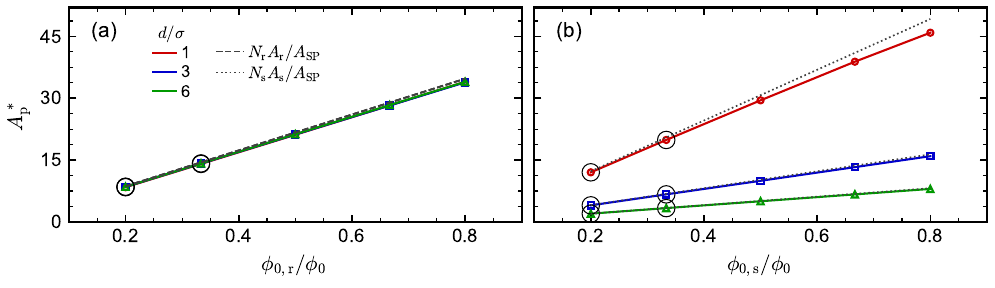}
    \caption{Normalized surface area of pores $A_{\rm p}^*$ after removing (a) spheres or (b) rods as a function of initial composition of the particle that remains, (a) rods or (b) spheres, for different sphere diameters $d$ when ${\rm Pe} = 5$. The black circles indicate cases where the remaining particle network was below the chosen percolation threshold, and the dashed lines are the total normalized surface area of the remaining particles. The estimated uncertainty is less than the marker size.}
    \label{fig:Surfacearea}
\end{figure*}

To quantify the additional surface area created by the pores, we normalized $A_{\rm p}$ by the surface area of a solid SP, i.e., $A_{\rm p}^* = A/A_{\rm SP}$ with $A_{\rm SP} = 4\pi R^2$. The normalized pore surface area $A_{\rm p}^*$ for ${\rm Pe} = 5$ is plotted in Figure \ref{fig:Surfacearea}; the corresponding results for ${\rm Pe} = 10$ and ${\rm Pe} = 50$ showed the same trends (Figures S12 and S13, respectively). For SPs obtained after removal of spheres, $A_{\rm p}^*$ increased linearly with increasing $\phi_{0, {\rm r}}/\phi_0$ and was virtually independent of sphere diameter $d$ (Figure \ref{fig:Surfacearea}a). These trends indicate that the pore surface area is controlled primarily by the surface are of the remaining rods, $N_{\rm r}A_{\rm r}$ where $A_{\rm r} = \pi\sigma^2 \left( 1+\lambda \right)$ is the surface area of a rod estimated as that of a spherocylinder. This behavior is consistent with the pore phase being almost fully connected in all cases (Figure S11a). For SPs obtained after removal of rods, an analogous picture emerged (Figure \ref{fig:Surfacearea}b): $A_{\rm p}^*$ increased with increasing $\phi_{0, {\rm s}}/\phi_0$, reflecting the larger pore-material interface provided by the increasing amount of remaining sphere material. In this case, however, $A_{\rm p}^*$ increased with decreasing sphere diameter $d$ at fixed sphere volume fraction, reflecting the larger surface-to-volume ratio of small spheres and the highly contiguous nature of the pore volume (Figure S11b). As a consequence, $A_{\rm p}^*$ closely followed $N_{\rm s}A_{\rm s}$ for each case, where $A_{\rm s} = \pi d^2$ is the surface area of a sphere.

\begin{figure*}
    \centering
    \includegraphics[width=\linewidth]{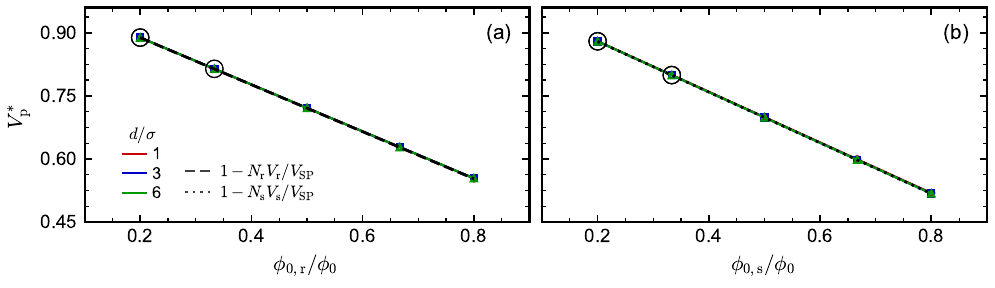}
    \caption{Normalized total volume of pores $V_{\rm p}^* = V_{\rm p}/V_{\rm SP}$ after removing (a) spheres or (b) rods as a function of initial composition of the particle that remains, (a) rods or (b) spheres, for different sphere diameters $d$ when ${\rm Pe} = 5$. The black circles indicate cases where the remaining particle network was below the chosen percolation threshold, and the dashed lines are the total normalized pore volume expected from the total volume of the particles removed. The estimated uncertainty is less than the marker size.}
    \label{fig:volumepore}
\end{figure*}

We also considered the pore volume normalized by that of a solid SP, $V_{\rm p}^*=V_{\rm p}/V_{\rm SP}$ where $V_{\rm SP} = 4\pi R^3/3$. For all cases, $V_{\rm p}^*$ decreased with increasing volume fraction of the remaining component, irrespective of whether spheres or rods were removed as well as the drying rate (Figure \ref{fig:volumepore} for ${\rm Pe} = 5$, Figure S14 for ${\rm Pe} = 10$, and Figure S15 for ${\rm Pe} = 50$). This behavior is expected because the total pore volume is largely determined by the amount of removed material; consistently, the measured pore volume was approximately $V_{\rm p} = V_{\rm SP} - N_i v_i$ where $i$ is the component that remains. Thus, unlike the pore surface area, $V_{\rm p}^*$ mainly reflects the amount of remaining material within the SP, which is insensitive to the size or shape of the removed content.

The volume and surface area of the pores are important properties of the SP, but the distribution of pore sizes also plays an important role for the transport of species through the SP. Accordingly, we characterized the distribution of local pore diameter $d_{\rm p}$ using the method developed by Gelb and Gubbins \cite{gelb:1999, bhattacharya:2006}. We inserted $10^5$ test points at random positions within the SP boundary but not inside a remaining particle and, for each point, determined the largest sphere that contained the insertion point without intersecting any remaining particles. Because each $d_{\rm p}$ is defined by a spherical probe that must contain its insertion point, off-center positions within a cavity left by a removed particle generally yield diameters smaller than the sphere diameter $d$. To avoid identifying the exterior of the SP surface as pores, the region beyond $51.5\,\sigma$ was filled with particles of diameter $\sigma$ arranged on a face-centered cubic lattice. The resulting probability distribution $p(d_{\rm p})$ characterizes the local pore sizes within the SP.

\begin{figure}
    \centering
    \includegraphics{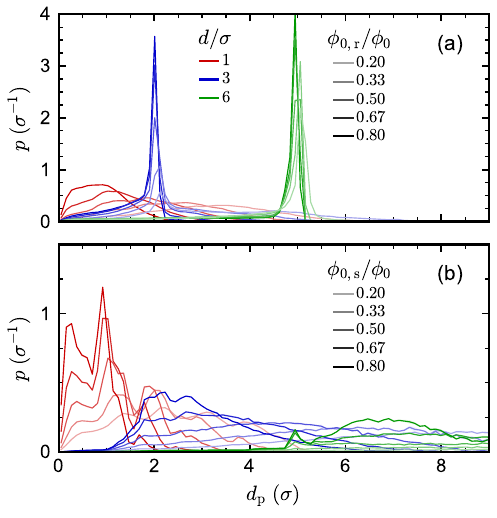}
    \caption{Probability density $p$ for pore diameter $d_{\rm p}$ after removal of (a) spheres or (b) rods for different sphere diameters $d$ and initial compositions when ${\rm Pe} = 5$.}
    \label{fig:poresizedistribution}
\end{figure}

Removing spheres resulted in relatively narrow distributions of $p(d_{\rm p})$ for $d=3\,\sigma$ and $d=6\,\sigma$ (Figure \ref{fig:poresizedistribution}a for ${\rm Pe}=5$, Figure S16a for ${\rm Pe} = 10$, and Figure S17a for ${\rm Pe} = 50$), reflecting the more isolated placement of these larger spheres within the dried SP (Figure \ref{fig:snapshots}). In contrast, the distribution was considerably broader for $d=1\,\sigma$, consistent with the greater variability in the clustering of small spheres prior to their removal. In all cases, $p(d_{\rm p})$ shifted toward large diameters with decreasing rod fraction $\phi_{0,{\rm r}}/\phi_0$. This shift indicates that sacrificial spheres increasingly came into close proximity before removal, so that neighboring voids merged into larger pores. This effect became more pronounced for smaller spheres because decreasing $d$ at fixed volume fraction increases the number of spheres and thus the probability that neighboring sacrificial particles merge into larger pore regions upon removal. When the rods acted as the sacrificial component (Figure \ref{fig:poresizedistribution}b for ${\rm Pe} = 5$, Figure S16b for ${\rm Pe} = 10$, and Figure S17b for ${\rm Pe} = 50$), the pore-size distributions were broader than in the sphere-removal case, indicating a wider range of local pore sizes. The distributions shifted toward larger $d_{\rm p}$ with increasing sphere diameter and decreasing sphere fraction.

\begin{figure*}
    \centering
    \includegraphics[width=\linewidth]{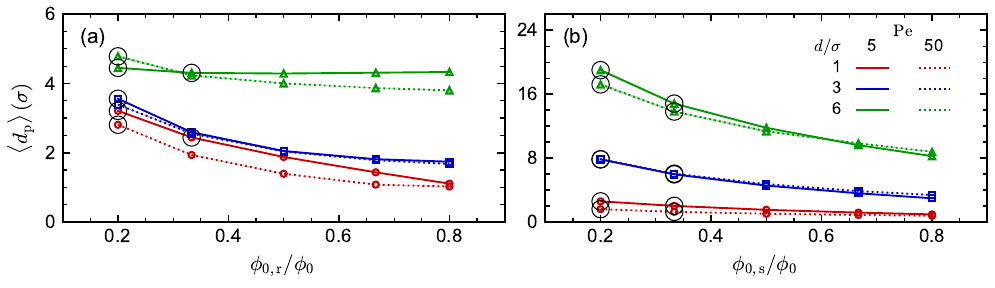}
    \caption{Mean pore diameter $\langle d_{\rm p} \rangle$ after removing (a) spheres or (b) rods as a function of initial composition of the particle that remains, (a) rods or (b) spheres, for different sphere diameters $d$ when ${\rm Pe} = 5$ or 50. The black circles indicate cases where the remaining particle network was below the chosen percolation threshold. See Figure S18 for the corresponding ${\rm Pe} = 10$ data, also compared to the ${\rm Pe} = 5$ data. The estimated uncertainty is less than the marker size.}
    \label{fig:psd}
\end{figure*}

Finally, we computed the mean pore diameter $\langle d_{\rm p}\rangle$ from $p\left( d_{\rm p} \right)$, which effectively characterizes the local cross-section of the pores. When the spheres were removed, $\langle d_{\rm p}\rangle$ had a nontrivial dependence on the size of the spheres (Figure \ref{fig:psd}a). For $d=1\,\sigma$, we found that $\langle d_{\rm p}\rangle$ roughly tripled from $\langle d_{\rm p}\rangle \approx \sigma$ to $\approx 3\,\sigma$ as the volume fraction of remaining rods decreased from $\phi_{0,{\rm r}}/\phi_0=0.80$ to $0.20$. Further, $\langle d_{\rm p}\rangle$ was slightly larger for SPs dried at ${\rm Pe}=5$ than at ${\rm Pe} = 50$, which we attribute to the formation of small nematic rod bundles and of sphere clusters at slow drying, resulting in larger voids after their removal. This trend was not observed at $\phi_{0,{\rm r}}/\phi_0=0.80$, because the number of spheres was too low to form appreciable clusters. Increasing the sphere diameter to $d=3\,\sigma$ caused only a modest further increase in $\langle d_{\rm p}\rangle$, which remained below $d$ for almost all systems. This small increase reflected differences in the shape of $p\left( d_{\rm p} \right)$: the distributions for $d=3\,\sigma$ showed sharper peaks, whereas those for $d=1\,\sigma$ were broader in some cases. Further, $\langle d_{\rm p}\rangle$ was nearly identical for the two drying speeds investigated here. For $d=6\,\sigma$, the mean diameter of the pores $\langle d_{\rm p}\rangle$ approached but remained below that of the removed spheres. In contrast to the two smaller sphere sizes, decreasing $\phi_{0, {\rm r}}/\phi_0$ increased $\langle d_{\rm p}\rangle$ only slightly at fast drying (${\rm Pe} = 50$) and had almost no effect at slow drying (${\rm Pe} = 5$), consistent with the relatively small number of large, isolated spheres in the dried SPs (Figure \ref{fig:snapshots}) and the correspondingly narrow $p(d_{\rm p})$ distributions (Figure \ref{fig:poresizedistribution}a).

When the rods were removed, $\langle d_{\rm p}\rangle$ generally decreased with increasing volume fraction of remaining spheres $\phi_{0, {\rm s}}/\phi_0$ (Figure \ref{fig:psd}b). This reduction was weak for $d=1\,\sigma$, with $\langle d_{\rm p}\rangle$ on the order of $\sigma$ for all compositions, indicating that the rods were well separated before removal and therefore generated pores comparable to the diameter of a single rod. For larger spheres, $\langle d_{\rm p}\rangle$ was larger and more composition-dependent, indicating the formation of local rod bundles before removal. Compared to sphere removal, the effect of drying speed was less pronounced. Overall, the mean diameter of the pores obtained by removing rods were generally larger than those obtained by removing spheres, reflecting the anisotropic shape and local bundling of the rod-shaped sacrificial component.

\section{Conclusions}
\label{section:conclusions}
We have examined how particle shape, initial composition, and drying influence the mesostructure of SPs formed from mixtures of rod-shaped and spherical particles with one component selectively removed to make pores. The balance between diffusion and advection in the drying process played an important role in setting the particle organization within the SP. Drying at higher Pe led to accumulation of rods at the interface. We also observed the formation of central voids when the SPs were formed at higher Pe. This effect also depended on the amount of spheres within the SP.

These differences in structure were reflected in the resulting pore properties. The pore-size distributions depended strongly on the shape and size of the sacrificial particles. Removing larger spheres produced larger pores, while increasing the amount of removed particles also shifted the pore-size distribution toward larger pore sizes. However, increasing the amount of sacrificial particles also reduced the amount of material left behind, and beyond a certain point the remaining particles no longer formed a percolating network. Rods retained connectivity down to lower remaining volume fractions than spheres because their elongated shape allowed them to form connected networks more readily. The pore surface area was mainly set by the exposed surface area of the particles that remained after removal. For spheres, this was controlled only by the amount of remaining rods, whereas for rod removal it also depended on sphere diameter because smaller spheres have a larger total surface area at fixed volume fraction. In contrast, the pore volume was mainly controlled by the amount of material removed and was nearly independent of particle size and drying rate. Taken together, these results show that the different properties of the pore morphology can be tuned by varying the particle shape and size as well as their initial composition and which is removed. These findings may hence have important implications for engineering the mechanical and functional properties of SPs.

\section*{Acknowledgements}
This material is based upon work supported by the National Science Foundation under Award No.~2223084. In addition, this work was supported by the Deutsche Forschungsgemeinschaft (DFG, German Research Foundation) through the framework of the research training group RTG 2767 (Project No. 451785257), and through Projects 470113688 and 509039598. This work was completed with resources provided by the National High Performance Computing Center of the Dresden University of Technology.

\section*{Supporting Information}
The following files are available free of charge.
\begin{itemize}
  \item Supporting Information: Additional simulation parameters; full set of volume fraction profiles and images of dried SPs; particle connectivity analysis for $\mathrm{Pe} = 10$; additional pore analysis; and details of voxel-based surface-area calculation.
\end{itemize}

\section*{Author contributions}
\noindent
{\bf Kritika Kritika:} Conceptualization (equal); Data curation (lead); Formal analysis (lead); Investigation (lead); Methodology (equal); Software (lead); Validation (equal); Visualization (lead); Writing - Original draft (lead); Writing - review \& editing (supporting)
\\
{\bf Michael P. Howard:} Conceptualization (equal); Formal analysis (supporting); Funding acquisition (equal); Investigation (supporting); Methodology (equal); Project administration (equal); Supervision (equal); Validation (equal); Writing - Original draft (supporting); Writing - review \& editing (equal)
\\
{\bf Arash Nikoubashman:} Conceptualization (equal); Formal analysis (supporting); Funding acquisition (equal); Investigation (supporting); Methodology (equal); Project administration (equal); Resources (lead); Supervision (equal); Validation (equal); Writing - Original draft (supporting); Writing - review \& editing (equal)

\printbibliography

\end{document}


\maketitle

\clearpage
\section{Simulation Parameters}
\begin{table}[H]
\centering
\caption{Number of rods $N_{\rm r}$ and number of spheres $N_{\rm s}$ for different sphere diameters $d$ and initial composition ratios $\phi_{0,\rm s}/\phi_{0, \rm r}$.}
\begin{tabular}{cccc}
\toprule
$d/\sigma$ & $\phi_{0, {\rm s}}/\phi_{0, {\rm r}}$ & $N_{\rm r}$ & $N_{\rm s}$ \\
\midrule
1 & 0.25 & 60235 & 128000 \\
1 & 0.5  & 50196 & 213333 \\
1 & 1    & 37647 & 320000 \\
1 & 2    & 25098 & 426666 \\
1 & 4    & 15058 & 512000 \\
\midrule
3 & 0.25 & 60235 &  4740 \\
3 & 0.5  & 50196 &  7901 \\
3 & 1    & 37647 & 11851 \\
3 & 2    & 25098 & 15802 \\
3 & 4    & 15058 & 18962 \\
\midrule
6 & 0.25 & 60235 &  592 \\
6 & 0.5  & 50196 &  987 \\
6 & 1    & 37647 & 1481 \\
6 & 2    & 25098 & 1975 \\
6 & 4    & 15058 & 2370 \\
\bottomrule
\end{tabular}
\end{table}

\begin{table}[H]
\centering
\caption{P\'{e}clet numbers for rods ${\rm Pe}_{\rm r}$ with corresponding P\'{e}clet numbers for spheres ${\rm Pe}_{\rm s}$ with diameter $d$ indicated in parentheses.}
\begin{tabular}{cccc}
\toprule
${\rm Pe}_{\rm r}$ & ${\rm Pe}_{\rm s}$ ($1\,\sigma$) & ${\rm Pe}_{\rm s}$ ($3\,\sigma$) & ${\rm Pe}_{\rm s}$ ($6\,\sigma$) \\
\midrule
5.0  & 1.84 & 5.52 & 11.03 \\
10.0 & 3.68 & 11.03 & 22.06 \\
50.0 & 18.38 & 55.16 & 110.31 \\
\bottomrule
\end{tabular}
\end{table}

\clearpage
\FloatBarrier
\section{Local Volume Fraction and Snapshots}
\begin{figure}[H]
    \centering
    \includegraphics[width=\linewidth]{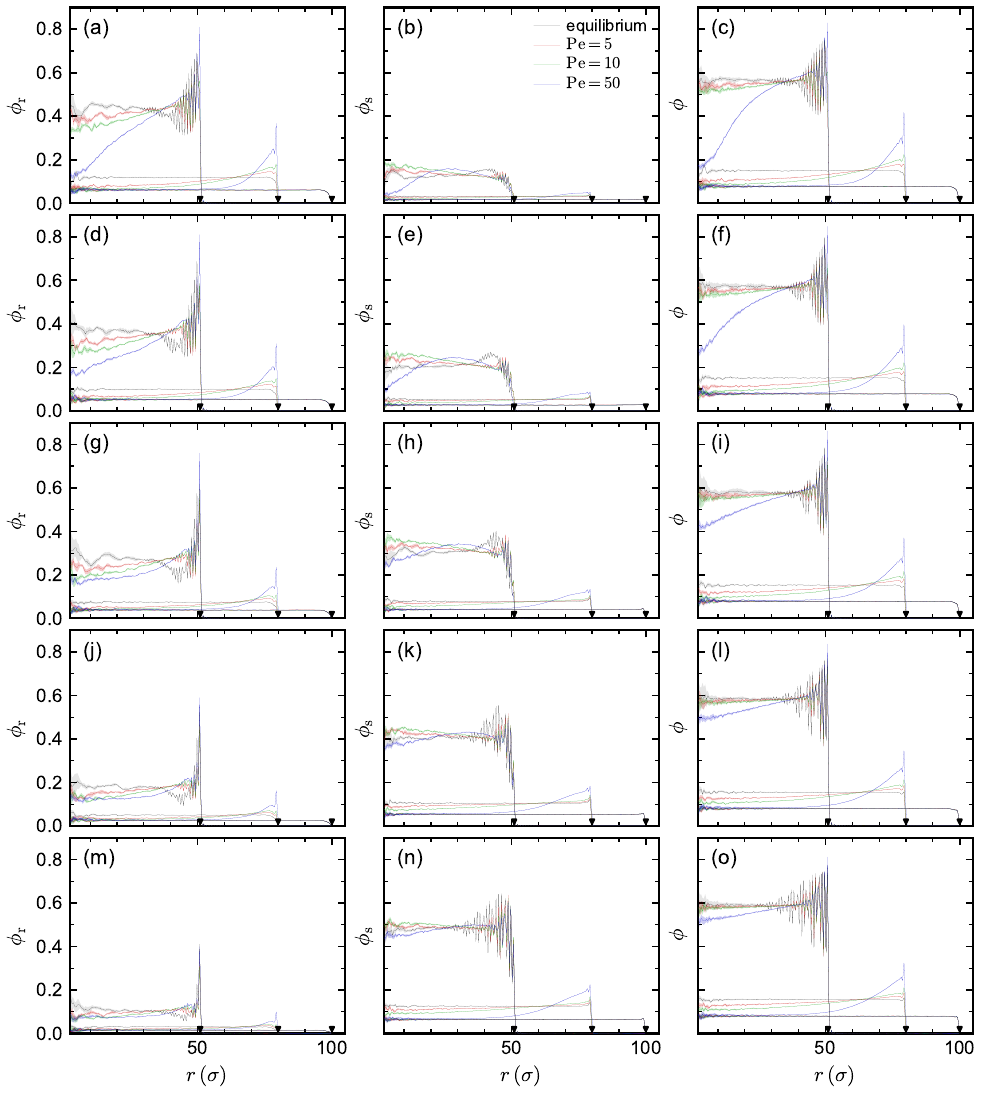}
    \caption{Local volume fraction of rods (a, d, g, j, m) $\phi_{\rm r}$, (b, e, h, k, n) spheres $\phi_{\rm s}$, and (c, f, i, l, o) their sum $\phi = \phi_{\rm r} + \phi_{\rm s}$ at radial distance $r$ from the center of the droplet for sphere diameter $d=1\,\sigma$ and different initial compositions $\phi_{0,{\rm s}}/\phi_0$: (a--c) 0.20, (d--f) 0.33, (g--i) 0.50, (j--l) 0.67, and (m--o) 0.80. Solid lines are profiles in the drying simulations initially ($R = R_0=100\,\sigma$), when the radius was about $R = 79.9\,\sigma$, and at the end ($R=51\,\sigma$) when ${\rm Pe} = 5$, 10, or 50, while dashed lines are the corresponding equilibrium profiles for these droplet radii.}
    \label{fig:S1}
\end{figure}

\begin{figure}[H]
    \centering
    \includegraphics[width=\linewidth]{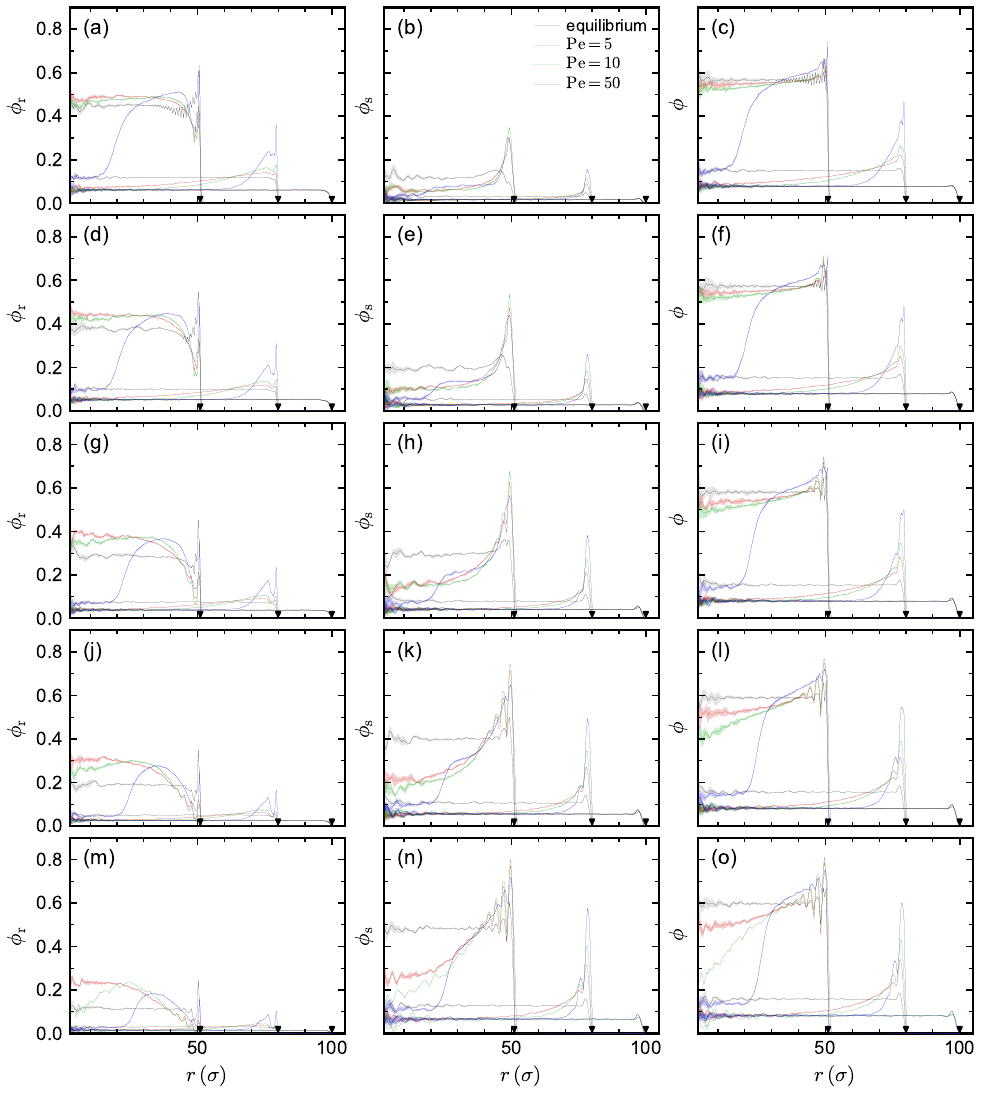}
    \caption{The same as Figure \ref{fig:S1} but for $d=3\,\sigma$.}
\end{figure}

\begin{figure}[H]
    \centering
    \includegraphics[width=\linewidth]{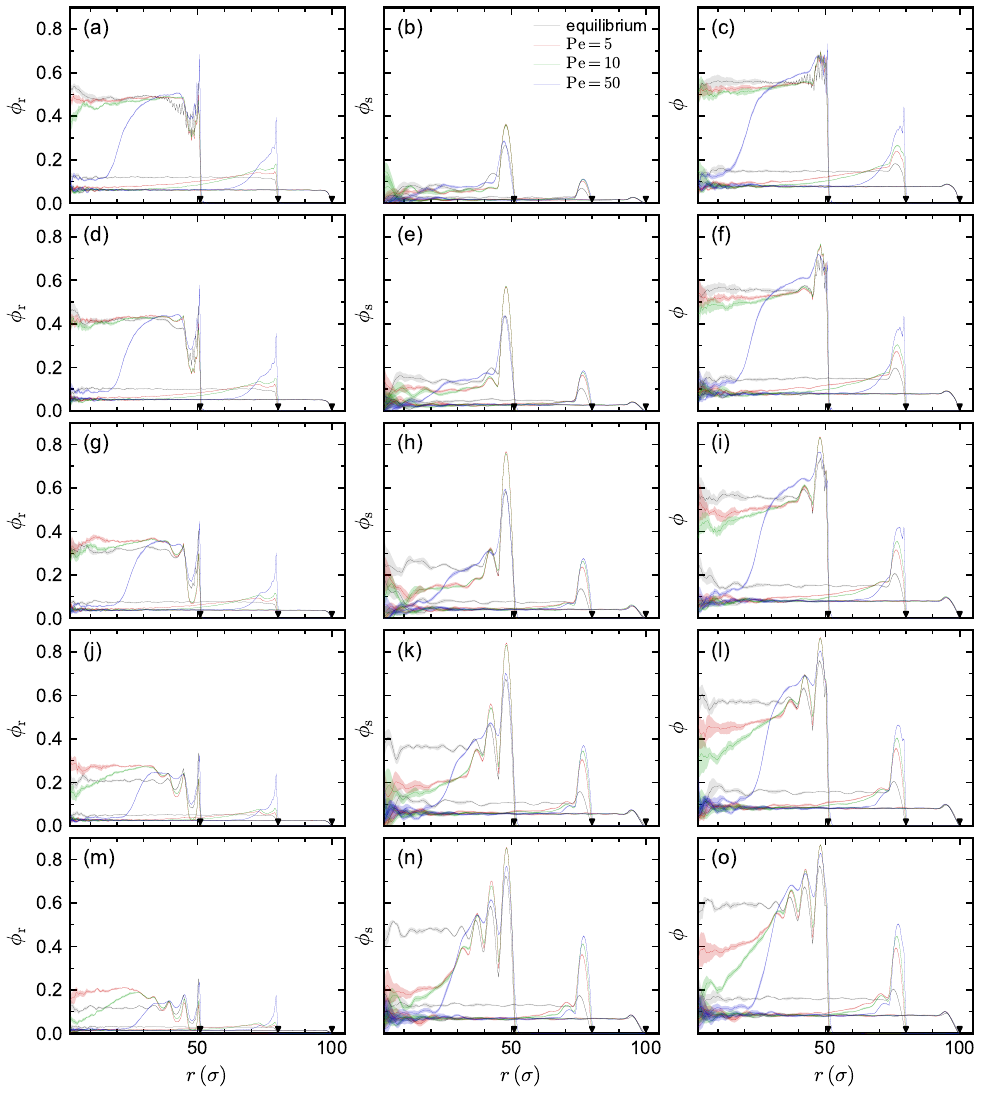}
    \caption{The same as Figure \ref{fig:S1} but for $d=6\,\sigma$.}
\end{figure}

\begin{figure}[H]
    \centering
    \includegraphics[width=\linewidth]{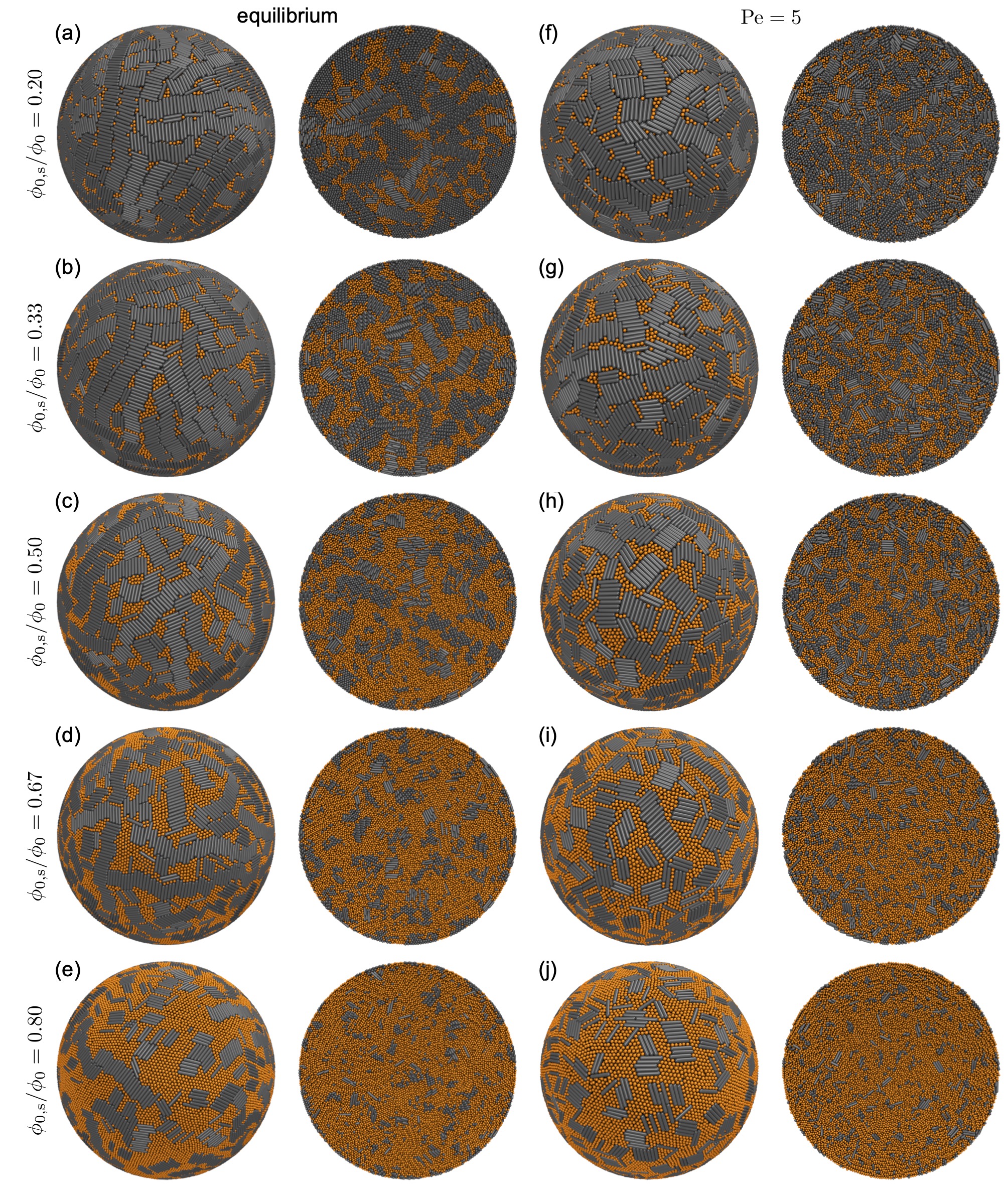}
    \caption{Exterior (left) and cross-sectional (right) views of SPs (a--e) in equilibrium and (f--h) when ${\rm Pe}=5$ for $d=1\,\sigma$ and varying $\phi_{0,{\rm s}}/\phi_0$: (a, f) 0.20, (b, g) 0.33, (c, h) 0.50, (d, i) 0.67, and (e, j) 0.80.}
    \label{fig:S4}
\end{figure}

\begin{figure}[H]
    \centering
    \includegraphics[width=\linewidth]{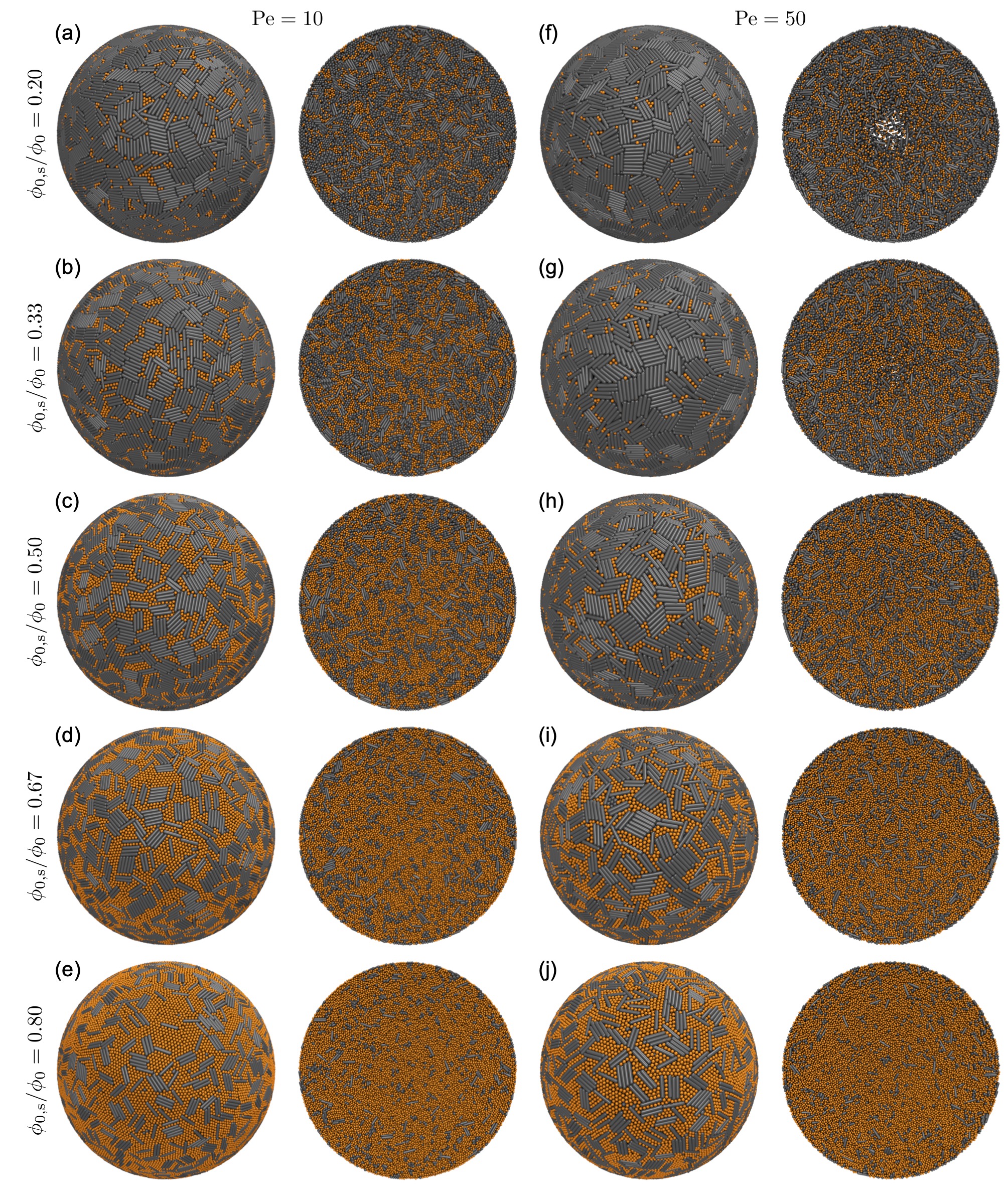}
    \caption{The same as Figure \ref{fig:S4} but with (a--e) showing ${\rm Pe}=10$ and (f--j) showing ${\rm Pe}=50$.}
    \label{fig:S5}
\end{figure}

\begin{figure}[H]
    \centering
    \includegraphics[width=\linewidth]{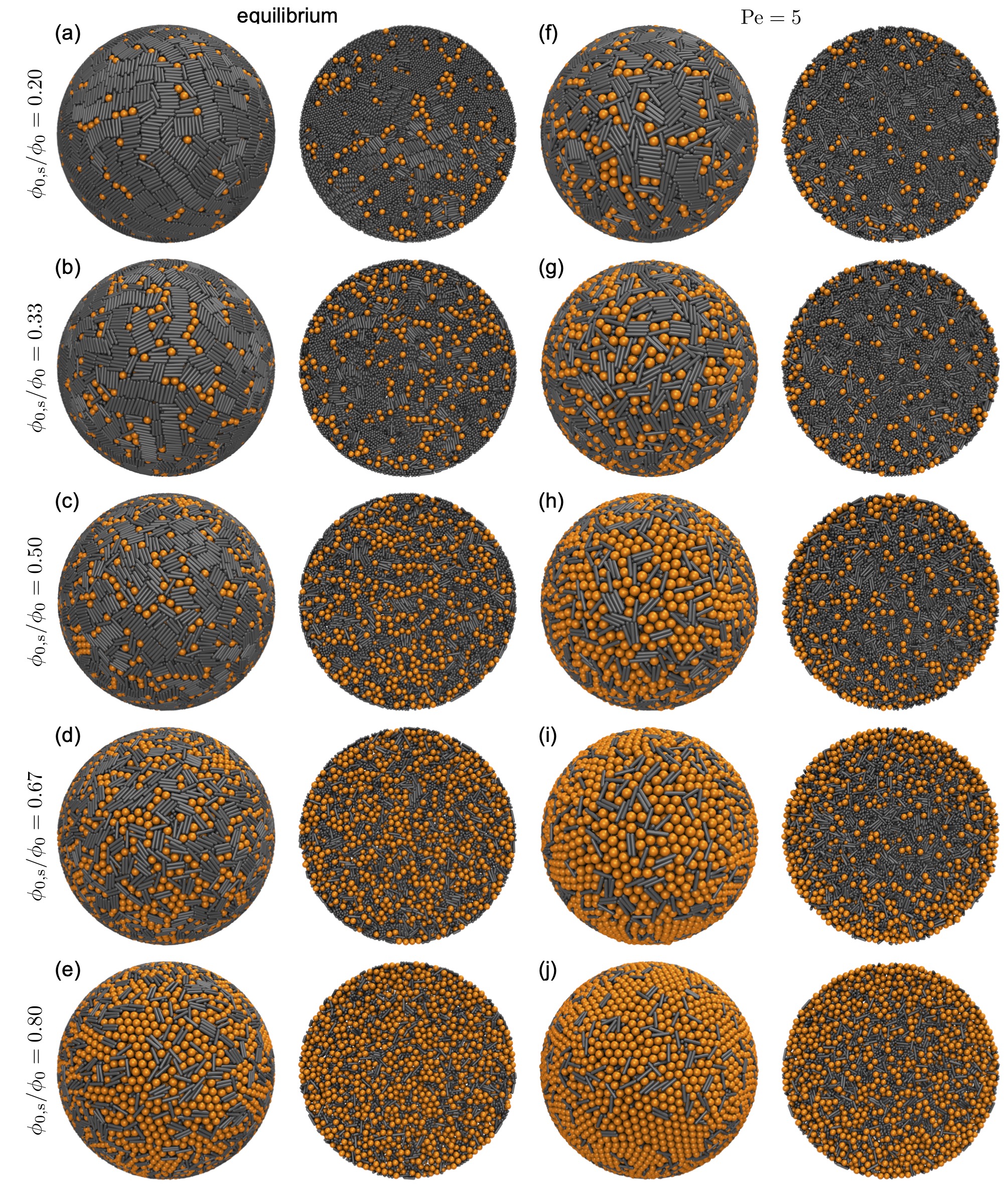}
    \caption{The same as Figure \ref{fig:S4} but for $d=3\,\sigma$.}
\end{figure}

\begin{figure}[H]
    \centering
    \includegraphics[width=\linewidth]{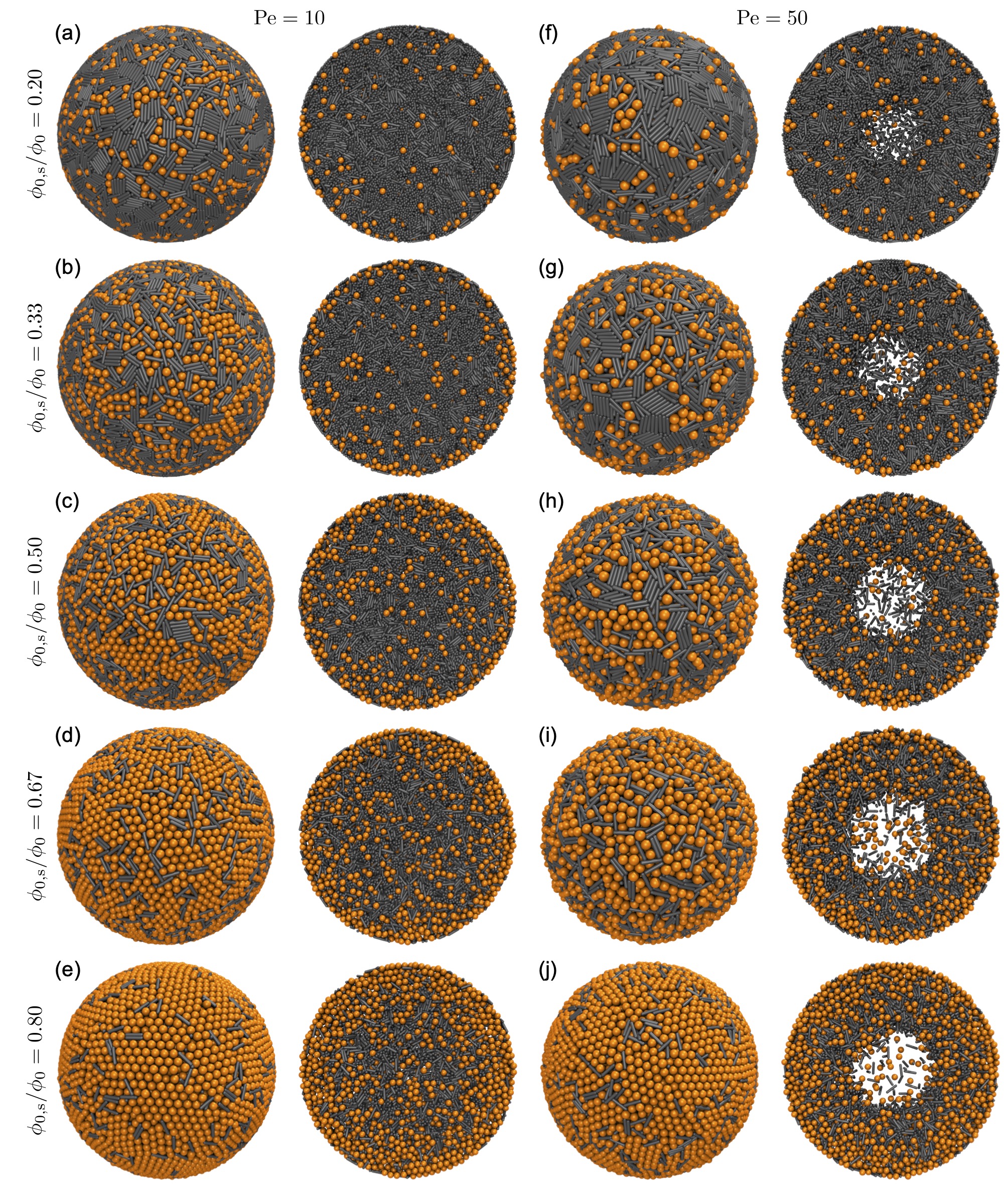}
    \caption{The same as Figure \ref{fig:S5} but for $d=3\,\sigma$.}
\end{figure}

\begin{figure}[H]
    \centering
    \includegraphics[width=\linewidth]{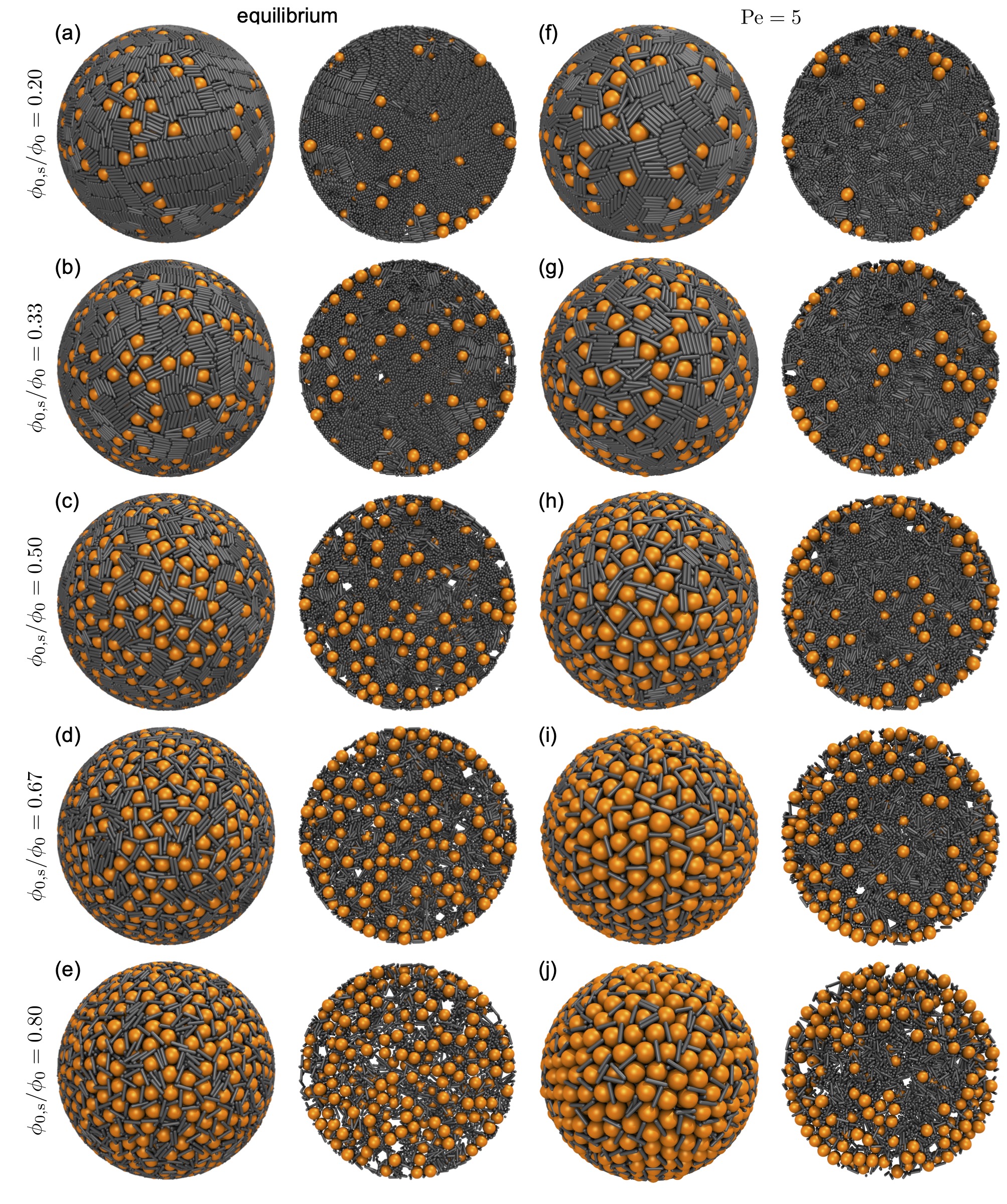}
    \caption{The same as Figure \ref{fig:S4} but for $d=6\,\sigma$.}
\end{figure}

\begin{figure}[H]
    \centering
    \includegraphics[width=\linewidth]{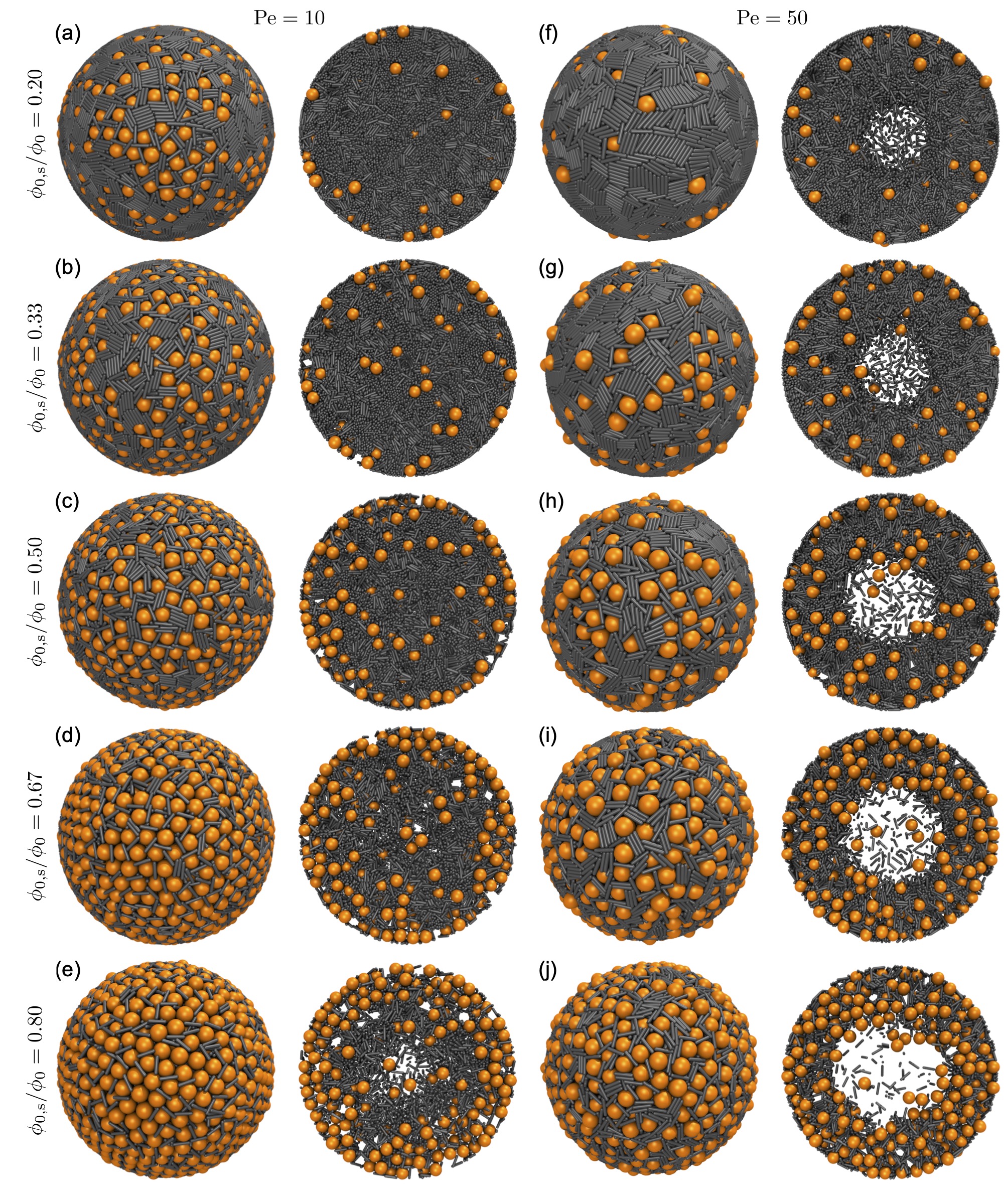}
    \caption{The same as Figure \ref{fig:S5} but for $d=6\,\sigma$.}
\end{figure}

\section{Particle Connectivity}
\begin{figure}[H]
    \centering
    \includegraphics[width=\linewidth]{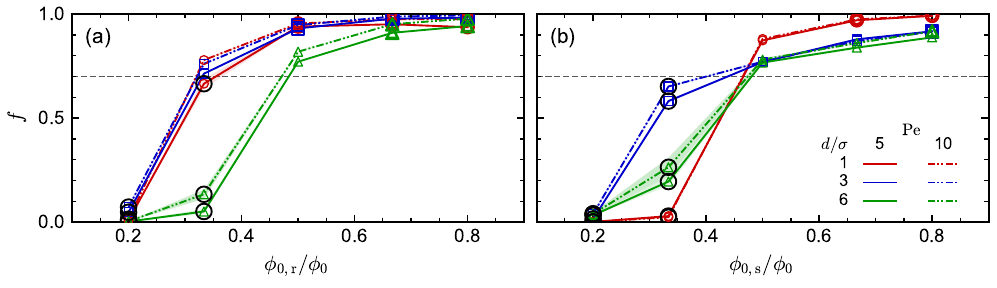}
    \caption{The same as Figure 5 but for ${\rm Pe}=10$ instead of ${\rm Pe}=50$.}
\end{figure}

\FloatBarrier
\section{Additional Pore Analysis}
\begin{figure}[H]
    \centering
    \includegraphics[width=\linewidth]{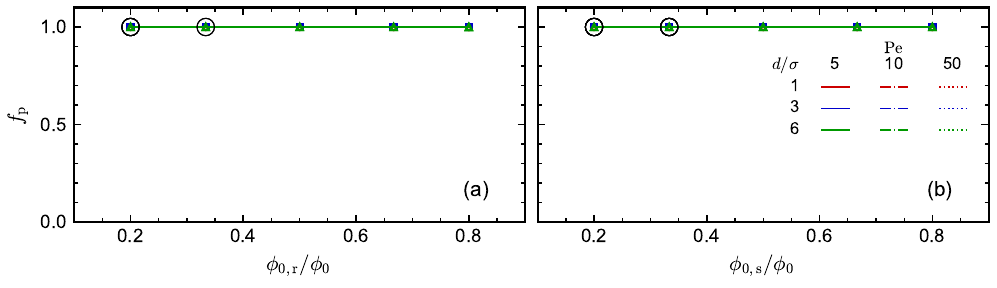}
    \caption{Fraction of pore voxels $f_{\rm p}$ in the largest cluster after removing (a) spheres or (b) rods as a function of initial composition of the particle that remains, (a) rods or (b) spheres, for all diameters, initial compositions, and Pe. The estimated uncertainty is less than the marker size.}
\end{figure}

\begin{figure}
    \centering
    \includegraphics[width=\linewidth]{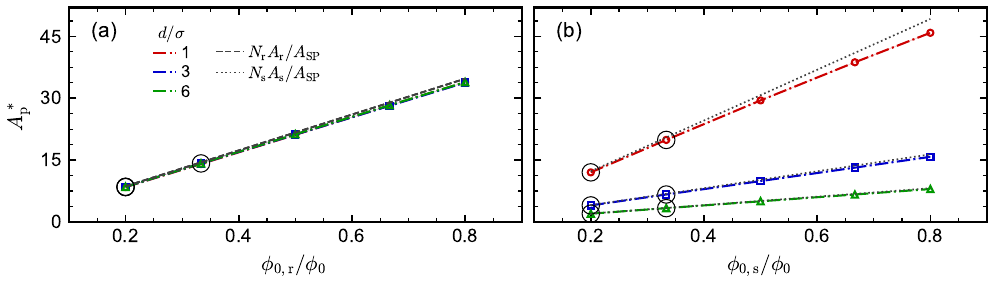}
    \caption{The same as Figure 6 but for ${\rm Pe} = 10$.}
\end{figure}

\begin{figure}
    \centering
    \includegraphics[width=\linewidth]{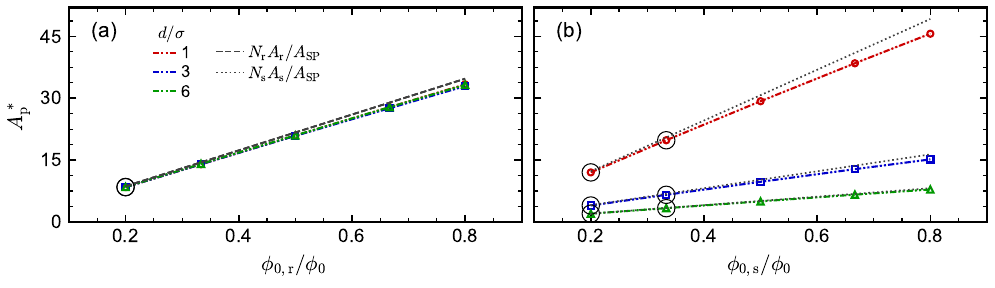}
    \caption{The same as Figure 6 but for ${\rm Pe} = 50$.}
\end{figure}

\begin{figure}
    \centering
    \includegraphics[width=\linewidth]{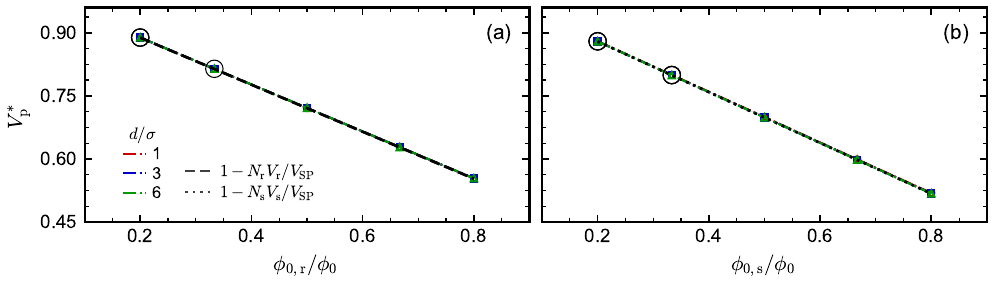}
    \caption{The same as Figure 7 but for ${\rm Pe}=10$.}
\end{figure}

\begin{figure}
    \centering
    \includegraphics[width=\linewidth]{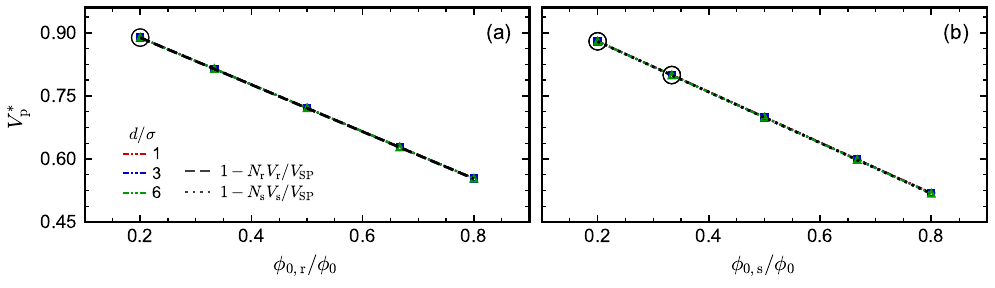}
    \caption{The same as Figure 7 but for ${\rm Pe} = 50$.}
\end{figure}

\begin{figure}[H]
    \centering
    \includegraphics[width=8cm]{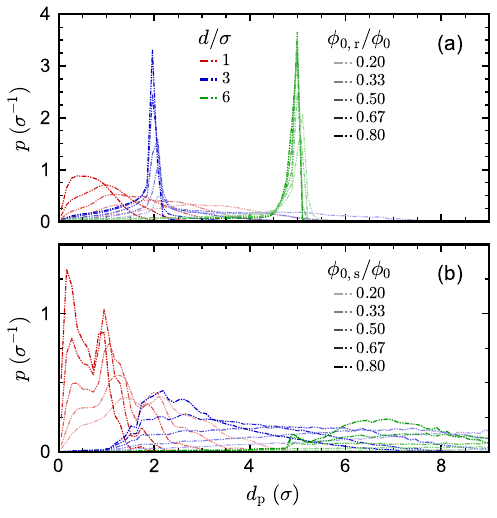}
    \caption{The same as Figure 8 but for ${\rm Pe} = 10$.}
\end{figure}

\begin{figure}[H]
    \centering
    \includegraphics[width=8cm]{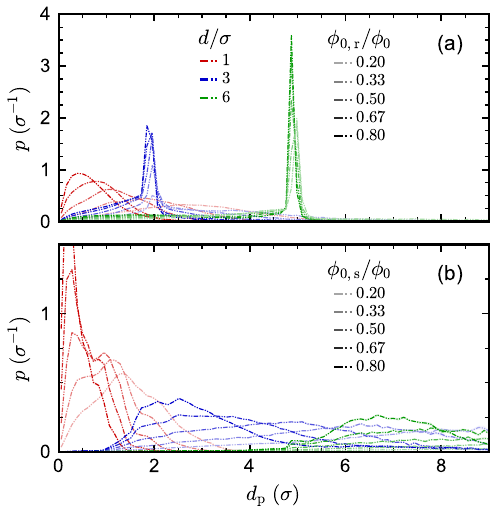}
    \caption{The same as Figure 8 but for ${\rm Pe} = 50$.}
\end{figure}

\begin{figure}
    \centering
    \includegraphics[width=\linewidth]{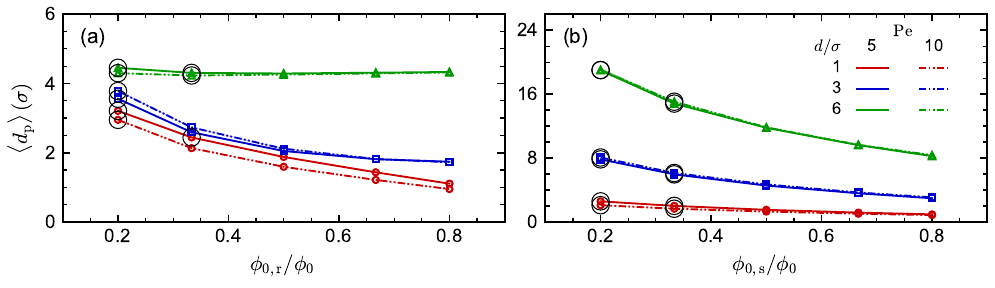}
    \caption{The same as Figure 9 but for ${\rm Pe}=10$ instead of ${\rm Pe}=50$.}
\end{figure}

\FloatBarrier
\section{Surface-Area Correction for Voxelized Interfaces}
The pore surface area was computed from the voxelized pores by counting exposed faces. This procedure systematically overestimates the true interfacial area because a smooth interface is represented by axis-aligned voxel faces. The voxel-face area associated with a smooth surface element ${\rm d}A$ can be understood from the projections of that element onto the three coordinate planes. For a surface element with unit normal $\mathbf{\hat{n}}=(n_x,n_y,n_z)$, the projected area on the $xy$ plane is $|n_z|{\rm d}A$. Similarly, the projections onto the $yz$ and $xz$ planes are $|n_x|{\rm d}A$ and $|n_y|{\rm d}A$, respectively. The exposed voxel-face area associated with this surface element is therefore approximated as the sum of these three projected areas, $(|n_x|+|n_y|+|n_z|){\rm d}A$. For a spherical interface, the surface normals are uniformly distributed over the unit sphere. By symmetry, $\langle |n_x|+|n_y|+|n_z| \rangle = 3\langle |n_z| \rangle$. Using $n_z=\cos\theta$ and ${\rm d}\Omega=\sin\theta{\rm d}\theta{\rm d}\phi$, we obtain
\begin{equation}
    \langle |n_z| \rangle = \frac{1}{4\pi}\int_0^{2\pi}\int_0^\pi |\cos\theta|\sin\theta{\rm d}\theta{\rm d}\phi = \frac{1}{2} .
\end{equation}
Therefore, for a smooth spherical interface, counting the exposed voxel-face area overestimates the actual surface area by a factor of $\langle |n_x|+|n_y|+|n_z| \rangle = 3/2$. Therefore, we multiplied the voxelized pore surface areas with a correction factor of 2/3. No correction is required for the voxelized pore volume.